\documentclass[aps,10pt,twocolumn,amsmath,amssymb,prl,superscriptaddress, floatfix]{revtex4-1}

\usepackage{amssymb,amsmath,amsfonts}
\usepackage{graphicx,color}
\usepackage[english]{babel}
\usepackage{tabularx}

\newcommand{\1}{\begin{equation}}
\newcommand{\2}{\end{equation}}
\newcommand{\ea}{\begin{eqnarray}} 
\newcommand{\ee}{\end{eqnarray}}
\newcommand{\4}[2]{{\frac{#1}{#2}}}
 
\newcommand{\Sum}[2]{{\sum\limits_{#1}^{#2}}}

\newcommand{\I}{{ {\rm i}  }}

\begin{document}

\title{Viscotaxis: microswimmer navigation in viscosity gradients}

\author{Benno Liebchen} 
\email{liebchen@hhu.de}
\affiliation{Institut f\"{u}r Theoretische Physik II: Weiche Materie, Heinrich-Heine-Universit\"{a}t D\"{u}sseldorf, D-40225 D\"{u}sseldorf, Germany}

\author{Paul Monderkamp} 
\affiliation{Institut f\"{u}r Theoretische Physik II: Weiche Materie, Heinrich-Heine-Universit\"{a}t D\"{u}sseldorf, D-40225 D\"{u}sseldorf, Germany}

\author{\surname{Borge} ten Hagen}
\affiliation{Physics of Fluids Group and Max Planck Center Twente, Department of Science and Technology,
MESA+ Institute, and J. M. Burgers Centre for Fluid Dynamics, University of Twente, 7500 AE Enschede,
The Netherlands}

\author{Hartmut L\"{o}wen}
\affiliation{Institut f\"{u}r Theoretische Physik II: Weiche Materie, Heinrich-Heine-Universit\"{a}t D\"{u}sseldorf, D-40225 D\"{u}sseldorf, Germany}

\date{\today}

\begin{abstract}
The survival of many microorganisms, like \textit{Leptospira} or \textit{Spiroplasma} bacteria, can depend on their ability to navigate towards regions of favorable viscosity.
While this ability, called viscotaxis, has been observed in several bacterial experiments, the underlying mechanism remains unclear. 
Here, we provide a framework to study viscotaxis of self-propelled swimmers in slowly varying viscosity fields
and show that suitable body shapes create viscotaxis based on a systematic asymmetry of viscous forces acting on a microswimmer.
Our results shed new light on viscotaxis in \textit{Spiroplasma} and \textit{Leptospira} and suggest that dynamic body shape changes exhibited by both types of microorganisms may have an unrecognized 
functionality: to prevent them from drifting to low viscosity regions where they swim poorly. 
The present theory 
classifies microswimmers regarding their ability to show viscotaxis 
and can be used to design synthetic viscotactic swimmers, e.g.\ for delivering drugs to a target region distinguished by viscosity.
\end{abstract}

\maketitle

\paragraph*{Introduction.}
\label{sec:intro}
The ability to adapt their motion in response to gradients in an external stimulus, called taxis \cite{Dusenbery}, is crucial for the life of 
most microorganisms. 
Chemotaxis \cite{Eisenbach}, for example, allows microorganisms to find food (chemoattraction or positive chemotaxis) 
and to escape from toxins (chemorepulsion or negative chemotaxis) but also acts as a precondition of mammalian life by guiding sperm cells towards the egg. Very recently,
it has been found that chemotaxis even plays an important role \cite{Hong2007,Pohl2014,Saha2014,Liebchen2015,Liebchen2016,Liebchen2017,Jin2017,Illien2017}
in synthetic colloidal microswimmers \cite{Romanczuk2012,Elgeti2015,Bechinger2016}, where it can induce dynamic clusters and patterns even at low density
\cite{Pohl2014,Saha2014,Liebchen2015,Liebchen2016,Liebchen2017}.
Besides responding to chemical stimuli, many biological and synthetic swimmers can adapt their motion also in response to other stimuli such as
gradients in light intensity (phototaxis) \cite{Jekely2008,Bennett2015,Giometto2015,Lozano2016,Dai2016}, magnetic fields (magnetotaxis) \cite{Klumpp2016,Rupprecht2016,Waisbord2016},
temperature (thermotaxis) \cite{Bahat2003,Li2013,Bregulla2016,Bickel2014}, or a gravitational potential (gravitaxis) \cite{Richter2007,Roberts2010,tenHagen2014,Campbell2013,Campbell2017,Wolff2013}. 
\\In this paper we consider ``viscotaxis'', which is a different kind of taxis caused by viscosity gradients. Viscotaxis is much less understood than the above types of taxes for microorganisms 
and undiscovered for synthetic swimmers. However, viscosity gradients occur in many situations,
both for biological and synthetic swimmers. The interface between 
two immiscible liquids of different viscosity, for example, features a high viscosity gradient, occurring, e.g., in a 
sedimentation profile of the two liquids. 
Similarly, viscosity gradients show up naturally in complex environments, e.g. in fluids near a mucus zone or at the interface of different parts of the human body with individual viscosities. 
Therefore, some microorganisms, like \textit{Spiroplasma} \cite{Daniels1980} or \textit{Leptospira interrogans} \cite{Kaiser1975,Petrino1978,Takabe2017}, which are poor swimmers at low viscosity \cite{Kaiser1975,Sherman1982}, 
have developed the ability to navigate up viscosity gradients. Without this ability to perform viscotactic motion, they would statistically migrate down the gradient \cite{Sherman1982} 
as the residence time of a particle 
in a certain volume element decreases with its speed. 
Despite the existence of a series of experiments on bacterial viscotaxis \cite{Petrino1978,Daniels1980,Sherman1982,Takabe2017},
there is no systematic theory nor an understanding of the precise mechanism 
allowing microorganisms to perform viscotaxis \cite{Takabe2017}.
We therefore develop a framework to study viscotaxis for self-propelled biological and synthetic microswimmers 
in slowly varying viscosity fields.
The present approach shows that \emph{nonuniaxial} body shapes \emph{automatically} lead to viscous torques aligning linear swimmers \emph{generically} up viscosity gradients. This is due to a systematic mismatch of viscous forces acting on 
different body parts of the swimmer.
The proposed mechanism may help to explain experiments observing positive viscotaxis in \textit{Spiroplasma} \cite{Daniels1980} and \textit{Leptospira} \cite{Kaiser1975,Petrino1978,Takabe2017}, 
which has previously been attributed to the speculative existence of viscoreceptors \cite{Sherman1982}.
The provided theory may also help linking the characteristic motility mode in \textit{Spiroplasma} and \textit{Leptospira}, which  
involves uniaxial and nonuniaxial body shapes \cite{Davis1973,Daniels1980,Gilad2003,Shaevitz2005,Takabe2017}, with the very fact that these organisms show viscotaxis
\cite{Petrino1978,Takabe2017}. 
We also demonstrate that swimmers experiencing nonviscous torques, like chiral swimmers or run-and-tumble bacteria,
can in principle generate negative viscotaxis, which may shed new light on corresponding observations for \textit{E.\ coli} bacteria \cite{Sherman1982}.
More generally, our theory classifies swimmers, based on their body shape and self-produced torques, regarding their ability to perform viscotaxis. 
This classification can be used 
as a new design principle for 
synthetic microswimmers that are able to navigate in viscosity gradients.
\begin{figure}
\includegraphics[width=\columnwidth]{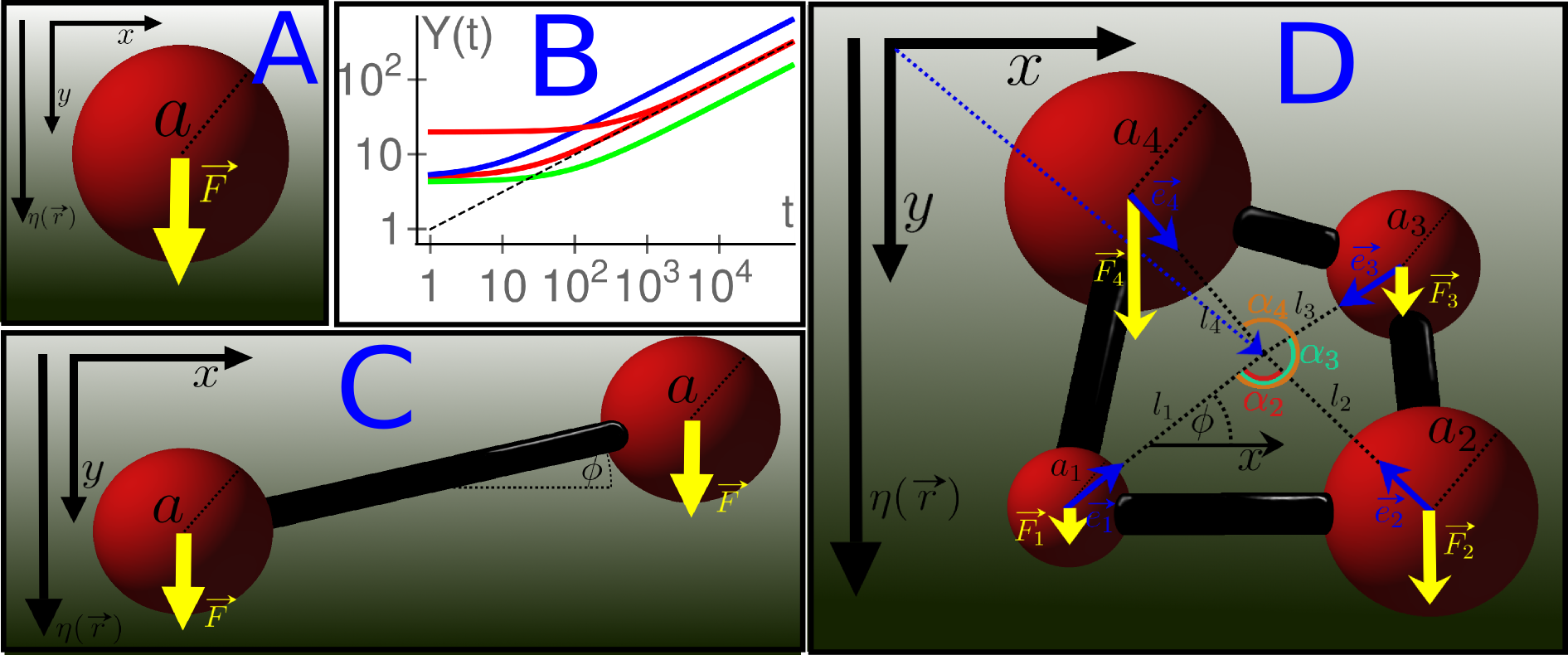}
\caption{The sedimentation of a sphere (A), a dumbbell (C), and more complex objects (D) 
in a linear viscosity profile $\eta(y)=\eta_0+\lambda y$ 
(illustrated by the shading of the background color)
universally follows a $Y(t) \propto \sqrt{t}$ law at late times (B), 
where $Y$ is the y component of the hydrodynamic center of mass (long blue arrow in D).
Panel B shows the sedimentation dynamics based on numerical integration of (\ref{eomf}, \ref{eomm})
of a single sphere (two red curves corresponding to two different initial $Y$-values); a symmetric dumbbell (blue) 
and an irregular triangle (green); (arbitrary parameters).   
The black dashed line shows the $y(t)=\sqrt{t}$ line for comparison.}
\label{fig1}
\end{figure}

\paragraph*{Passive particles in viscosity gradients.}
To develop a first understanding for the dynamics in viscosity gradients, we consider a passive and  
overdamped rigid body which is confined to two dimensions (quasi 2D) and consists 
of $N$ rigidly connected spheres with midpoints ${\bf r}_i=(x_i,y_i)$ ($i=1..N$) and radii $b_i$. 
Besides external forces ${\bf F}_i$, which we will later replace with self-propulsion `forces' \cite{tenHagen2015,Samin2015,Takatori2016}, the spheres also experience viscous 
forces. Since we are mainly interested in slowly varying viscosity fields,
in the sense that viscous forces do not change much on the scale of a single sphere [$|\eta(y_i+b_i)-\eta(y_i)|/\eta(y_i) \ll 1$],
we apply Stokes's law to define the frictional forces ${\bf F}_R({\bf r}_i)\approx -6 \pi \eta({\bf r}_i) b_i \dot{\bf r}_i$ (for effects due to large viscosity gradients see \cite{Malgaretti2016}). 
This leads us to the
following equations of motion
\ea 
&&\sum \left[ {\bf F}_i- 6\pi b_i \eta({\bf r}_i) \dot {\bf r}_i \right] = \mathbf{0}, \label{eomf}\\
&&\sum ({\bf r}_i-{\bf R}) \times \left[ {\bf F}_i -6\pi b_i \eta({\bf r}_i) \dot {\bf r}_i\right] = 0, \label{eomm}
\ee
which can be seen as conditions that the swimmer has to be force- and torque-free.
Here, ${\bf a} \times {\bf b}= a_1 b_2-a_2 b_1$, $\sum:=\Sum{i=1}{N}$, and ${\bf R}=(X,Y)$ is some reference point for which we choose, for convenience, 
the hydrodynamic center of mass ${\bf R}:=\sum b_i {\bf r}_i/\sum b_i$;
characteristically, forces acting at this point do not cause rotations of the rigid body in the lab frame. 
We focus here on linear viscosity profiles $\eta({\bf r})=\eta_0+\lambda y$, and eliminate $\eta_0$ by shifting the origin of the coordinate system $y\rightarrow y -
\eta_0/\lambda$ and $\lambda$ by defining a rescaled radius $a_i=6\pi \lambda b_i$.
\\To get a first intuition for the dynamics of particles in viscosity gradients, we consider externally 
forced bodies, say, the 
`sedimentation' of a sphere 
experiencing a constant force ${\bf F}_1= F_0{\bf e}_y$ pointing up the gradient (Fig.~\ref{fig1}A). 
Will the sphere move forever or stop at large viscosity? 
For a single sphere, Eqs.~(\ref{eomf}, \ref{eomm}) 
yield $y_1(t)=y_1(t=0)+\sqrt{2F_0 t/a_1}$ and $x_1(t)=x_1(t=0)$ showing that the sphere continuously slows down as $\dot y_1 \propto 1/\sqrt{t}$, but never stops moving. 
\\While more complicated $N$-body objects experiencing forces ${\bf F}_i= F_i{\bf e}_y$ feature a more involved short-time dynamics, at late times, they universally follow the same
$\sqrt{t}$ law as the single sphere. 
To see this, we write, for later convenience, ${\bf r}_i={\bf R}-l_i {\bf e}_i$ (cf.\ Fig.~\ref{fig1}D), where $l_i=|{\bf r}_i-{\bf R}|$ and ${\bf e}_i=\left(\cos\phi_i,\sin \phi_i \right)$
with $\phi_i=:\phi+\alpha_i$ being the angle between ${\bf R}-{\bf r}_i$ and the $x$ axis and $\phi$ defining the orientation of the swimmer based on the orientation of ${\bf R}-{\bf r}_1$ relative to the $x$ axis (Figs.~\ref{fig1}C,D).
For $Y \gg l_i$, we have $y_i =Y-l_i \sin\phi_i \approx Y$ so that~(\ref{eomf}) 
yields
$(\sum a_i) Y\dot Y=\sum F_i$ resulting in $Y(t)=\sqrt{2 t (\sum F_i)/(\sum a_i)}$. 
Numerical simulations of (\ref{eomf}, \ref{eomm}) for a sphere, a symmetric dumbbell and a 
triangular swimmer confirm this law (see Fig.~\ref{fig1}B).
Note that nonlinear viscosity profiles lead to a qualitatively analogous behavior 
\footnote{Power-law profiles $\eta(y) \propto y^n$, for example, lead to
$Y^n \dot Y=({\rm d}/{\rm d}t) Y^{n+1}/(n+1)$ resulting in $Y(t)\propto t^{1/(n+1)}$; and exponential profiles $\eta(y)\propto {\rm e}^{\alpha y}$
lead to $Y(t) \propto \ln(t)$ resembling the slow-down in linear profiles.}.
\\Conversely to the position ${\bf R}(t)$, the late-time orientation of a sedimenting body in a viscosity gradient depends on its shape.  
To see this, let us consider a symmetric dumbbell with equal forces ${\bf F}= F{\bf e}_y$ acting on the two spheres (Fig.~\ref{fig1}C) and ask how it will align to the gradient 
while falling.
To find the answer, we use the late-time solution of (\ref{eomf}): $(X^\ast,Y^\ast)=\left(X(t=0),2\sqrt{t F/(a_1+a_2)}\right)$, 
and  
determine the fixpoints of (\ref{eomm}) as $\phi^\ast=0,\pi/2$, representing horizontal and vertical dumbbell orientations, respectively.
Performing a linear stability analysis of these fixpoints shows 
that the vertical configuration is always unstable,
whereas the horizontal one ($\phi^\ast=0$) attracts $\phi(t)$, which makes the dumbbell fall horizontally. 
Physically, for $\phi \neq 0$ the drag acting on the lower sphere (Fig.~\ref{fig1}C) dominates such that the upper sphere moves faster and the dumbbell turns 
towards the horizontal configuration (see also movie 1 in the Supplemental Material (SM) \cite{note_SM}).
\paragraph*{Microswimmer viscotaxis.}
We now turn to our key aim of exploring the mechanism allowing microswimmers to navigate in viscosity gradients.  
Analogously to the previous considerations for externally forced passive bodies, we describe active microswimmers as multi-bead rigid bodies \cite{Najafi2004,Earl2007,Pickl2012,Dunstan2012,Pande2015,Farutin2016,Vutukuri2017}; 
since our key results will 
apply to arbitrarily complex arrangements of (very small) beads, the model can also be used to closely mimic swimmers with a continuous shape.
For clarity and generality of the physical discussion, we implement activity by effective propulsion forces \cite{tenHagen2011,Takatori2016}, though our results largely 
apply also to force-free swimmers moving, e.g., by body deformations (see SM \cite{note_SM}). The latter type of swimmers experience only minor corrections from hydrodynamic interactions (SM \cite{note_SM}).
The effective forces point in fixed directions in the body frame of the particle and corotate with the swimmer in the laboratory frame. 
Formally, the dynamics of active swimmers therefore still follows Eqs.~(\ref{eomf}, \ref{eomm}), but with forces $\mathbf{F}_i\rightarrow \mathbf{F}_i(\phi)$
depending on the angle $\phi$ between the $x$ axis and the axis connecting sphere 1 and ${\bf R}$ (see Fig.~2B).
\begin{figure*}
\includegraphics[width=0.95\textwidth]{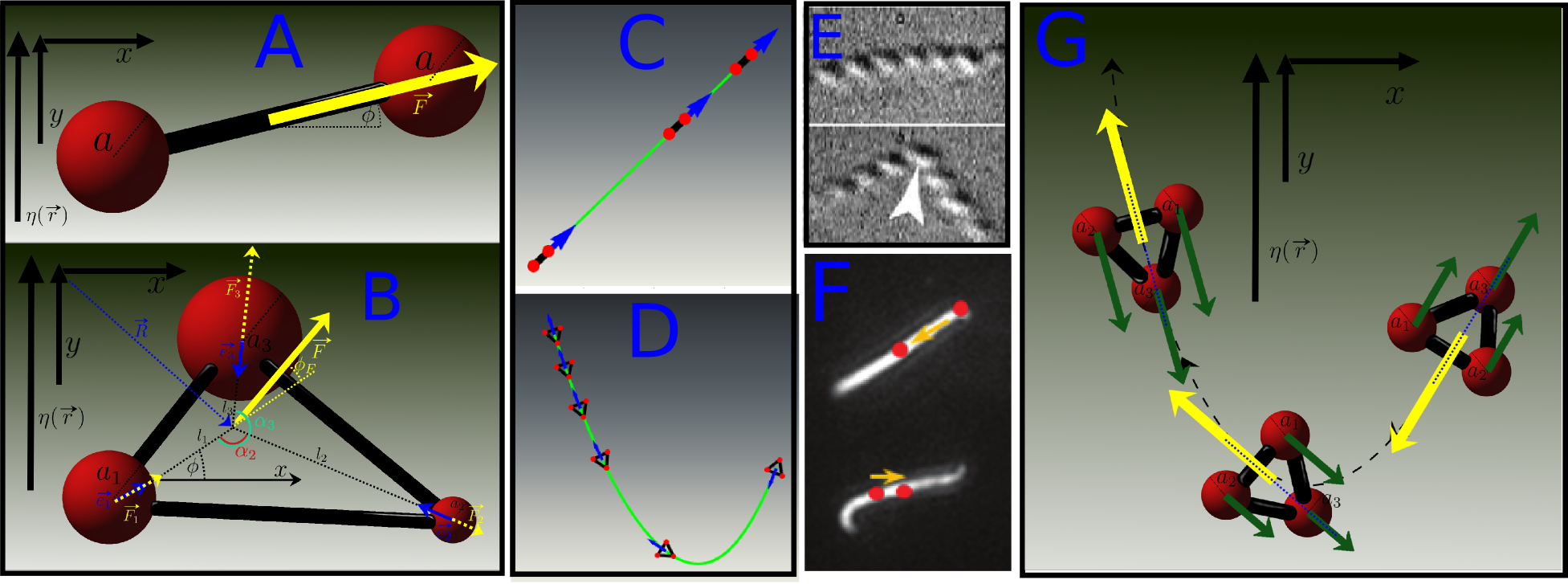}
\caption{Viscotaxis in linear swimmers:
Uniaxial swimmers (A) do not show viscotaxis and move in the direction of their initial orientation (exemplaric trajectory from numerical simulations in arbitrary units shown in C).  
Nonuniaxial swimmers (B) generically show viscotaxis which is always positive for linear swimmers (exemplaric trajectory in D). 
(E) \textit{Spiroplasma} and (F) \textit{Leptospira} bacteria changing their body shape from roughly uniaxial to nonuniaxial shapes (snapshots from \cite{Shaevitz2005,Takabe2017}). 
(G) Mechanism of viscotaxis: The viscous drag acting on body parts at high viscosity (sphere 1) dominates the drag acting on spheres at low viscosity (sphere 2), which turns the swimmer up the gradient.} \label{fig2}
\end{figure*}

\paragraph*{Minimal design of viscotactic swimmers.}
Let us focus on viscotaxis in linear swimmers and disregard chiral swimmers for now.  
Accordingly, we only allow for effective forces ${\bf F}_i \parallel ({\bf r}_i-{\bf R})=:{\bf x}_i$ 
\footnote{Forces not parallel to ${\bf r}_i-{\bf R}$ (i) produce torques which can generically 
not be canceled by viscous torques as the latter ones 
depend on the speed of the swimmer whose $\sqrt{t}$ time dependence in turn forbids a permanent canceling and (ii) cases where torques originating from different forces cancel each 
other do not create any new physics and can be disregarded.},
which sum up to a single force ${\bf F}$ pointing at some fixed angle $\phi_F$ relative to the swimmer (see Fig.~\ref{fig2}B), i.e.\
${\bf F}/|{\bf F}|=\left(\cos(\phi+\phi_F),\sin(\phi+\phi_F)\right)$. (Note that any net force acting on a point of the swimmer distinct from ${\bf R}$ would indeed create chirality.)
\\We now seek for the minimal design of a viscotactic swimmer.
Clearly, for a single sphere we have ${\bf r}_1={\bf R}$, such that (\ref{eomm}) is identically fulfilled and the sphere does not 
change its initial direction of motion. The next simplest candidate, a dumbbell
with a propulsion force pointing along the symmetry axis connecting both spheres (Fig.~\ref{fig2}A), experiences only viscous forces along its
symmetry axis but no viscous torque as ${\bf F} =  {\bf F}_1+{\bf F}_2$ and ${\bf x}_{1,2}$ all point along the symmetry axis of the dumbbell (compare Eq.~(\ref{eomm}) and Fig.~\ref{fig2}C).  
Following this argument, uniaxial swimmers cannot show viscotaxis (see also movie 2 in the SM \cite{note_SM}). (This result remains true in three dimensions of course and, as we show based on a perturbative solution 
of the Stokes equation in the presence of a viscosity gradient in the SM \cite{note_SM}, it is robust against 
hydrodynamic far-field interactions up to terms on the order of the relative change of viscosity on the scale of a single bead $a\lambda/\eta({\bf R})$.)
To see if triangular swimmers, as the simplest remaining candidate, can show viscotaxis, we now numerically solve Eqs.~(\ref{eomf}, \ref{eomm}) for a regular 
triangular swimmer (Fig.~\ref{fig2}B) with $a_1=a_2=a_3$, $l_1=l_2=l_3$, and $\phi_F=0$.
When the swimmer is initialized such that the propulsion forces push it up the viscosity gradient, it continues swimming in this direction. 
However, when we initialize it such that it starts swimming down the gradient, remarkably, we observe that the swimmer slowly turns and 
finally approaches a direction leading, again, to motion up the gradient (see Fig.~\ref{fig2}D and movies 3a,b in the SM).
Therefore, suitable body shapes can create viscotaxis. 
Figure~\ref{fig2}G illustrates the underlying mechanism: 
here, for a given $\dot {\bf R}$ sphere 1 experiences more drag than sphere 2 and moves more slowly ($f_1>f_2$ for the magnitudes of the friction forces). As a result the 
swimmer turns up the gradient until torque balance between the two spheres is reached. 
Once the swimmer has reached its late-time orientation, the active force points in a constant direction (vertically upwards for the regular triangular swimmer in Fig.~\ref{fig2}G) and we encounter the same universal $\sqrt{t}$ law as for passive particles in an external force field.
These arguments should apply analogously in three dimensions, of course. 
\\Repeating our simulations for other, less regular swimmers, we always find viscoattraction. 
To see how representative this result is for arbitrarily complicated swimmers, we next develop a systematic theory for viscotaxis.

\paragraph*{Theory of viscotaxis.} 
To understand viscotaxis in all types of microswimmers, we need to generally predict the late-time swimming direction $(X_0,Y_0)$ depending on the swimmer geometry. 
We first exploit the $\sqrt{t}$ scaling law and use the ansatz $X(t)= X_0 \sqrt{t}$, $Y(t)= Y_0 \sqrt{t}$ to asymptotically solve (\ref{eomf}, \ref{eomm}). 
As a result (see SM \cite{note_SM}), we find 
two possible late-time swimming directions determined by
\1
Y_0 = X_0 \tan\left(\phi_0+\phi_F\right), \label{tanexp}
\2
where $\phi_F$ is fixed by the propulsion forces (Fig.~\ref{fig2}B) and $\phi_0$ is given by
\1
\phi_0=\pm \arctan\left(\sum \left[c_i \cos(\alpha_i)\right]/\sqrt{\Sum{i,j=1}{N}c_i c_j \cos(\alpha_i-\alpha_j)}\right) \label{phifp}
\2
with $c_i=a_i l_i^2/(a_1 l_1^2)$ and the angles $\alpha_i$ (with $\alpha_1=0$) characterizing the swimmer geometry (see Fig.~\ref{fig1}D).
The two solutions in (\ref{phifp}) represent fixpoints of the orientational dynamics of the swimmer.
In fact, depending on system parameters, one or the other of these solutions accurately agrees with the late-time swimming direction seen in our simulations (not shown). 
To predict which of these solutions attracts the swimmer dynamics, we perform a linear stability analysis in the SM \cite{note_SM}.
The resulting growth rate of small orientational fluctuations around $\phi_0$ reads
\1 \sigma =\4{\sum c_i \sin\left(\phi_0-\phi_F + 2\alpha_i \right)}{2\sin(\phi_0+\phi_F)\sum c_i}. \label{growthrate}\2
A $\phi_0$ solution is stable if $\sigma<0$. 
The remaining task is to find some qualitatively informative relation between the sign of $\sigma$ in (\ref{growthrate}) and the swimmer geometry. 
Remarkably, it is possible to show \cite{note_SM} that $\sigma$ is generally negative if $\phi_0+\phi_F \in (0,\pi)$: that is, we universally find stability of the fixpoint (\ref{phifp})
representing motion (diagonally) up the viscosity gradient.
The conclusion is that, within our model, nonuniaxial linear swimmers generically move up viscosity gradients, no matter what their size and shape may be. 
\begin{figure}
\includegraphics[width=0.4\textwidth]{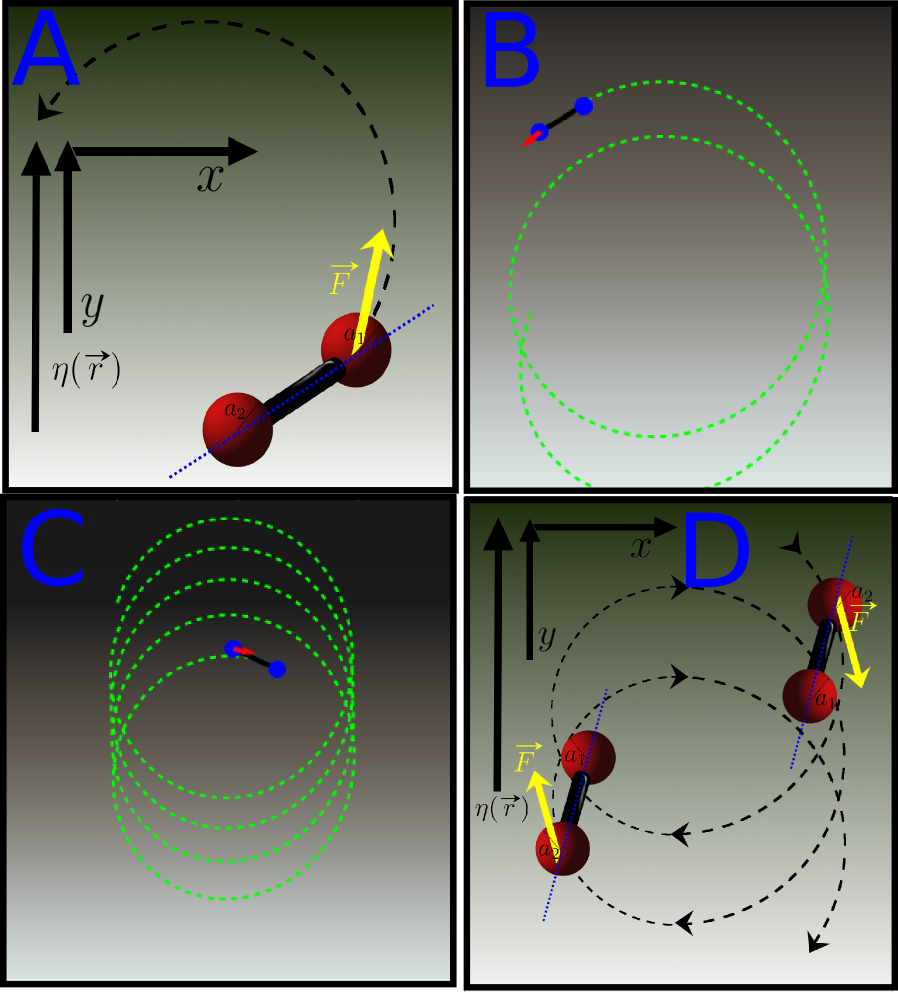}
\caption{Viscotaxis in chiral swimmers (A) can be positive (B, exemplaric trajectory from numerical simulations in arbitrary units) 
or negative (C). Negative viscotaxis emerges from the fact that back-driven swimmers (C,D) move faster down the gradient than upwards and systematically overshoot down the gradient while circling (see text for details).}
\label{fig3}
\end{figure}

\paragraph*{Biological implications.}
This conclusion is remarkable from a microbiological viewpoint. 
Both \textit{Leptospira interrogans} \cite{Kaiser1975, Petrino1978,Takabe2017} and \textit{Spiroplasma} bacteria \cite{Daniels1980} show viscoattraction based on a mechanism which is still unclear \cite{Takabe2017}.
Interestingly, the characteristic motility modes of both swimmer types contain conformational body changes involving sequences of uniaxial and nonuniaxial shapes; i.e., following the 
present theory, they should be automatically viscoattractive (rather than requiring speculative viscorecetors as it has been discussed previously; compare \cite{Sherman1982}). 
In fact, since both microswimmer types are poor swimmers in low-viscosity fluids, without viscotaxis they would statistically migrate to low viscosity regions \cite{Sherman1982}, where they would be quite inefficient 
in finding food.
In other words, the present theory suggests that
conformational body deformations in \textit{Leptospira} and \textit{Spiroplasma} may have a previously unrecognized functionality: to continuously align bacteria  
towards viscosity regions where they are efficient swimmers.

\paragraph*{Torque-induced negative viscotaxis.}
For nonlinear swimmers like circle swimmers or run-and-tumble bacteria, which experience a torque, 
negative viscotaxis is no longer forbidden by the above results. In fact, this class of swimmers can show both positive and negative viscotaxis.
To see this in detail, we reconsider our dumbbell-swimmer, but now allow the force to point at a finite angle to the swimmer axis (Fig.~\ref{fig3}A).
Numerically solving (\ref{eomf}, \ref{eomm}) for dumbbells swimming 
with the forcing sphere ahead, we generically observe a spiraling motion up the viscosity gradient (cf.\ Fig.~\ref{fig3}B and movie 4 in the SM), i.e. effectively viscoattraction.
Remarkably, however, dumbbells swimming with the forcing sphere in the back typically spiral 
down the gradient and effectively show viscorepulsion (cf.\ Fig.~\ref{fig3}C and movie 5 in the SM).
\\Physically, from the viewpoint of the forcing sphere, the second sphere provides more additional drag when the dumbbell moves up the viscosity gradient
than when moving down the gradient (Fig.~\ref{fig3}D). Therefore, the dumbbell swims faster down the gradient and overshoots in this direction within each circle.
The inverse situation applies to a dumbbell moving with the forcing sphere ahead (Fig.~\ref{fig3}B). 
\\Interestingly, some run-and-tumble bacteria also exhibit viscorepulsion \cite{Sherman1982}; these bacteria 
swim by means of flagella (forcing body part) in the back and show linear swimming periods (runs) which are interrupted by tumbles where the 
flagella create a torque randomizing the bacterial swimming direction.
Similarly to our dumbbell model, they should therefore swim more effectively during run sequences pointing down the gradient than during those pointing upwards, which could be responsible for their viscorepulsive behavior.

\paragraph*{Conclusions.}
Nonuniaxial body shapes create viscotaxis, which is generally positive in linear swimmers. 
This generic finding might explain the classical observations of positive viscotaxis seen in \textit{Spiroplasma} and \textit{Leptospira} bacteria;
in particular, it suggests that the transitions between uniaxial and nonuniaxial body shapes shown by both microorganisms,
previously attributed only to the propulsion mode of these microorganisms, may have an additional functionality:  
to prevent them from drifting towards regions where they swim poorly.
Here, viscotaxis emerges from a generic mechanism hinging on a systematic imbalance of viscous forces acting on the individual body parts of a swimmer. The same mechanism 
should apply to synthetic swimmers and may serve as a design principle e.g.\ for targeted drug delivery to regions of 
high viscosity.  

\section{Supplemental Material}

\subsection{Universality of viscoattraction for linear swimmers: fixpoint and linear stability analysis}
Here we calculate the fixpoints of the late-time dynamics of swimmers with arbitrary size and shape and analyze their linear stability.

We plug $X(t)= X_0 \sqrt{t}$, $Y(t)= Y_0 \sqrt{t}$ into Eq.~(\ref{eomf}) of the main text, 
where $(X_0,Y_0)$ determines the late-time swimming direction of the active body in the laboratory frame and $\phi(t)=\phi_0$ its late-time orientation (see Fig.~2B of the main text). 
To simplify the resulting expressions, we use the following identity which follows immediately from our
definition of the hydrodynamic center of mass (${\bf R}=\sum b_i {\bf r}_i/\sum b_i$)
\1 \Sum{i=1}{N} a_i l_i \cos(\phi_i)=0 \label{comrel} \2
and holds true for all geometrically possible combinations of $\phi_i=\phi+\alpha_i,a_i,l_i$.
Using this relation cancels most of the emerging terms and leads to 
the following special solutions of Eq.~(\ref{eomf}) in the main text:
$X_0 Y_0 \sum a_i=2F \cos(\phi_0 + \phi_F)$ and $Y_0^2 \sum a_i=2F \sin(\phi_0 + \phi_F)$, which can be combined to the useful relation
\1
Y_0 = X_0 \tan\left(\phi_0+\phi_F\right). \label{tanexpsm}
\2
This equation shows that the active swimmer moves, at late times, at an relative angle $\phi_0+\phi_F$ to the $x$ axis.
To find $\phi_0$ we now plug in $X(t)\rightarrow X_0 \sqrt{t}$, $Y(t)\rightarrow Y_0 \sqrt{t}$ 
along with (\ref{tanexpsm}) into Eq.~(\ref{eomm}) of the main text, which yields:
\1
\Sum{i=1}{N} a_i l_i^2\cos(\phi_0 +\phi_F) = \Sum{i=1}{N}a_i l_i^2\cos\left(\phi_0 - \phi_F +2\alpha_i \right). \label{fpb}
\2 
Summing up the trigonometric functions allows to straightforwardly solve (\ref{fpb}), yielding two possible late-time orientations (fixpoints of the orientational dynamics)
\1
\phi_0=\pm \arctan\left(\4{\Sum{i=1}{N}c_i \cos(\alpha_i)}{\sqrt{\Sum{i,j=1}{N}c_i c_j \cos(\alpha_i-\alpha_j)}}\right). \label{phifpsm}
\2
Here, the coefficients $c_i=a_i l_i^2/(a_1 l_1^2)$ and the angles $\alpha_i$ (with $\alpha_1=0$) are geometrically fixed (see Fig.~1D of the main text).
To see which $\phi_0$ solution acts as an attractor for the orientational dynamics, we now perform a linear stability analysis. 
Using $\left(X(t),Y(t),\phi(t)\right)=\left(X^\ast(t),Y^\ast(t),\phi^\ast(t)+\phi'(t)\right)$, where the $^\ast$ symbol indicates the late-time solution, and linearizing Eqs.~(\ref{eomf}, \ref{eomm}) (main text) in the fluctuations $\phi'$ yields by repeatedly employing (\ref{comrel}) 
$\phi'(t) = {\rm const}\cdot t^\sigma$ for the dynamics of the orientational fluctuation. Here, 
$\sigma$ is the growth rate of orientational fluctuations and given by 
\1 \sigma =\4{\sum c_i \sin\left(\phi_0-\phi_F + 2\alpha_i \right)}{2\sin(\phi_0+\phi_F)\sum c_i}. \label{growthratesm}\2
If $\sigma>0$ for a given parameter combination and $\phi_0$-solution, the corresponding fixpoint is unstable, and if $\sigma<0$, it is stable. 
The remaining challenge is now to rewrite (\ref{growthratesm}) such that it directly relates qualitative properties of the multi-bead swimmer with its late-time swimming direction $(X_0,Y_0)$, which is determined by $\phi_0$ and $\phi_F$.   
To achieve this, we first sum up harmonic functions again to write (\ref{growthratesm}) as
\1 \sigma = \4{A\sin\left(\phi_0-\phi_F + \delta \right)}{2\left(\sum c_i\right)\sin(\phi_0+\phi_F)},  \label{grate}\2
where $A>0$ (specific value is irrelevant) and 
\1
\delta = 2 \arctan\left(\4{\sum c_i \sin(2\alpha_i)}{\sum c_i^2 + \sum c_i \cos(2\alpha_i)}\right). \label{pshift}
\2
Now using specific information about the fixpoint by writing (\ref{fpb}) as
$\left(\sum c_i\right) \sin\left(\phi_0+\phi_F+\pi/2\right)=\sum c_i \sin\left(\phi_0-\phi_F + 2\alpha_i+\pi/2\right)=A\sin(\phi_0-\phi_F+\pi/2 +\delta)$ [same $A$ and $\delta$ as in (\ref{grate})]
and solving for $\delta$ allows us to write 
(\ref{grate}) in the useful form
\1
\sigma = -\4 {A \sqrt{1-x^2}}{2 \sin(\phi_0+\phi_F) \sum c_i},  \label{lambdares}
\2
where $x=(\sum c_i/A) \sin(\phi_0+\phi_F +\pi/2)$ which can be rewritten with the relation above (\ref{lambdares}) as $x=\sin(\phi_0-\phi_F+\pi/2+\delta)$ so that $|x|\leq 1$.
Following (\ref{lambdares}) ${\rm sign}[\sigma]={\rm sign}[-\sin(\phi_0+\phi_F)]$
such that, strikingly, the simple expression $\sin(\phi_0+\phi_F)$ determines the stability of the steady state solutions (\ref{phifpsm}) alone.
If $\phi_0+\phi_F \in (0,\pi)$, i.e. for upwards motion of the swimmer, we generally obtain $\sigma<0$, i.e. stability of the upwards pointing solution (viscoattractive).  
The conclusion is that nonuniaxial linear swimmers generically move up viscosity gradients (in the validity regime of our model), no matter what its size and shape may be. 
Small viscosity gradients therefore act as diodes for nonuniaxial active swimmers.

\subsection{Perturbation theory for the Stokes equation in a viscosity gradient}
To understand the impact of viscosity gradients on the hydrodynamic far-field interactions among individual beads, we now perturbatively solve the Stokes equation for a point
force in the presence of a viscosity gradient:
\ea
\nabla\cdot[\eta({\bf r})\nabla {\bf v}({\bf r})]-\nabla p({\bf r}) &=& -{\bf F}\delta({\bf r}_0), \label{stokes}\\
\nabla \cdot {\bf v}({\bf r}) &=& 0, \label{stokes2}
\ee
where ${\bf a}{\bf b}$ is the standard tensor product 
of vectors ${\bf a}$ and ${\bf b}$ and $\eta({\bf r})=\eta_0+\lambda(y-y_0)=\eta_0 \left[1+\epsilon \4{y-y_0}{L}\right]$ with $\eta_0:=\eta({\bf r}_0)$
and $\epsilon=\lambda L/\eta_0$ (we are interested in values of $y-y_0 \lesssim L$ and $\epsilon \ll 1$ so that $\eta({\bf r})$ is positive).
Here, $\epsilon$ measures the change in viscosity over the length scale of a swimmer compared to the viscosity of some part of the swimmer and serves as 
a suitable small dimensionless parameter for a perturbative solution of (\ref{stokes}).
Plugging the ansatz
\ea
{\bf v}&=&\Sum{n=0}{\infty}\epsilon^n {\bf v}_n,\\
p &=& \Sum{n=0}{\infty}\epsilon^n p_n
\ee 
into (\ref{stokes}) and equating terms of equal power in $\epsilon$ leads to the normal Oseen solution in order $\epsilon^0$ (with constant viscosity $\eta({\bf r}_0)$):
\ea
{\bf v}_0({\bf r}) &=& \4{I + \hat {\bf R}\hat {\bf R}}{8\pi \eta({\bf r}_0) R}\cdot {\bf F}, \\
p_0({\bf r}) &=& \4{\hat {\bf R}\cdot {\bf F}}{4\pi R^2}.
\ee
Here, $\hat {\bf R}={\bf R}/R$ with ${\bf R}={\bf r-r}_0$ and $R=|{\bf R}|$, and $I$ represents the 3$\times$3-identity matrix. 
The equation determining the $\mathcal{O}\left(\epsilon^1\right)$ term reads
\ea
\eta_0\nabla \cdot \left[\4{y-y_0}{L}\nabla {\bf v}_0\right] + \eta_0 \nabla^2 {\bf v}_1-\nabla p_1 &=& \mathbf{0}, \label{eq1}\\
\nabla \cdot {\bf v}_1 &=& 0. \label{eq1b}
\ee
Fourier transforming both equations (${\bf r}\rightarrow {\bf k}$), multiplying (\ref{eq1}) with ${\bf k}$, and using (\ref{eq1b}) we readily find 
\1
p_1({\bf k}) = -{\rm i} {\bf k} \cdot {\bf f}_0({\bf k}),
\2
where ${\bf f}_0({\bf k})$ is the Fourier transform of $(y-y_0) {\bf v}_0({\bf r})/L$.
Plugging this result for $p_1({\bf k})$ back into (\ref{eq1}) and transforming back to real space, we find 
\1
{\bf v}_1({\bf r})=-\left(\4{y-y_0}{L}\right){\bf v}_0({\bf r})+\4{1}{(2\pi)^3} \int \4{{\bf k}{\bf k}}{k^2} {\rm e}^{-\I {\bf k}\cdot {\bf x}}\cdot {\bf f}_0({\bf k}){\rm d}{\bf k}. \label{v1rpr}
\2
The only remaining challenge is to bring this expression into a useful form. To accomplish this, we (i) use the fundamental solution $u(r)=1/(4\pi r)$ of the Laplace equation 
$\nabla^2 u({\bf r})=-\delta({\bf r})$ to derive the Fourier 
transform of $\4{{\bf k}{\bf k}}{k^2}$ to configuration space  
\1 \4{1}{(2\pi)^3}\int \4{{\bf k}{\bf k}}{k^2}{\rm e}^{{\rm i} {\bf k}\cdot {\bf r}}  {\rm d}{\bf k} =\nabla \nabla \4{1}{4\pi r}\2
and (ii) apply the convolution theorem allowing us to rewrite (\ref{v1rpr}) as 
\ea
&&{\bf v}_1({\bf r}) = -\left(\4{y-y_0}{L}\right) {\bf v}_0({\bf r}) \nonumber \\
&+&\4{1}{4\pi}\int \4{y'-y_0}{L}  \left[\4{3\hat{\bf R}\hat {\bf R}}{R}-\4{I}{R}\right]_{{\bf R}={\bf r}-{\bf r}'} \cdot {\bf v}_0({\bf r}') {\rm d}{\bf r}'. \quad \label{gradcorr}
\ee
Notable conclusions from this result are: 
(i) The lowest correction to the far-field velocity profile induced by a point force in a medium of uniform viscosity [Oseen solution ${\bf v}_0({\bf r})$]
shows the same $1/r$ behavior as ${\bf v}_0({\bf r})$.
\\(ii) The coefficient of the correction term scales as
$\mathcal{O}\left(\4{\lambda L}{\eta({\bf r}_0)}\right)$.
Since hydrodynamic interactions at the level of the unperturbed Oseen tensor are of order $\mathcal{O}(a/L)$ 
(compare microswimmer model in the next section), small
viscosity gradients have an overall impact on the order of $\mathcal{O}\left(\4{\lambda a}{\eta({\bf r}_0)}\right)$ on the dynamics of 
a microswimmer. 
That is, they are negligible if the relative change of the viscosity 
on the order of a sphere radius is small -- an assumption that underlies the basic equations of our model anyway. 
Therefore hydrodynamic far-field interactions do not provide an additional constraint to the applicability of the present theory of viscotaxis.

\subsection{Microswimmer model}
Here, we develop a model of a microswimmer in a viscosity gradient which moves due to body shape deformations rather than due to effective propulsion forces.
Despite the presence of a viscosity gradient, our model is somewhat similar to the ``pushmepullyou swimmer'' \cite{Avron2005};
(a Najafi-Golestanian-like swimmer \cite{Najafi2004,Grosjean2016} or models considered in \cite{Pickl2012,Pande2015} could be used equally well.)
One key aim in this section is to show that the finding that uniaxial swimmers are incapable of showing viscotaxis survives for a swimmer which moves by body shape oscillations rather than by effective propulsion forces.

Consider a swimmer consisting of two spheres, (B) for body and (P) for propulsion, with coordinate vectors ${\bf R}(t)$ and ${\bf R}(t)+L(t){\bf e}(t)$ 
and radii $a_B$ and $a_P(t)$. Here, ${\bf e}(t)$
represents the unit vector pointing from (B) to (P) and $L(t)$ is the time-dependent length of the swimmer. 
We assume that the spheres are small in the sense that the viscosity does not change much over their size. We also assume that the distance between the 
spheres is large compared to their size so that far-field hydrodynamic interactions are sufficient to describe the swimmer dynamics. 

The drag forces acting on the spheres (B, P) then read
\ea
{\bf F}_B &=&-6\pi a_B \eta({\bf R})[\dot {\bf R}-{\bf u}({\bf R})], \\
{\bf F}_P &= & -6\pi a_P(t) \eta({\bf R}+L{\bf e})[\dot {\bf R}+\dot L{\bf e} + L\dot {\bf e}-{\bf u}({\bf R}+L{\bf e})], \nonumber
\ee
where ${\bf u}({\bf r})$ denotes the velocity of the surrounding fluid. 
The corresponding drag torques are
\ea
{\bf T}_B &=& ({\bf R}-{\bf R}_0) \times {\bf F}_B,\\
{\bf T}_P &=&  ({\bf R}+L{\bf e}-{\bf R}_0)\times {\bf F}_P,
\ee
where ${\bf R}_0$ is some reference vector, for which we should choose the hydrodynamic center of mass ${\bf R}_0=[(a_B+a_P(t)) {\bf R}+a_P(t)L(t){\bf e}(t)]/[a_B+a_P(t)]$.

The flow field at the position of (B) due to the force acting on the fluid at the position of (P) reads
\1
{\bf u}({\bf R}) =-G(-L{\bf e})\cdot {\bf F}_P
\2
and vice versa the flow field at the position of (P) due to the force acting on the fluid at the position of (B) reads
\1
{\bf u}({\bf R}+{\bf e}L) = -G(L{\bf e}) \cdot {\bf F}_B,
\2
where $G$ is the Green's function to (\ref{stokes}, \ref{stokes2}). The lowest order correction to the ``Oseen tensor'' due to viscosity gradients
follows immediately from (\ref{gradcorr}).

We now show that also in the framework of the present, hydrodynamically consistent model, uniaxial linear swimmers do not show viscotaxis: 
let us consider the ``dry'' case first, where ${\bf u}={\bf 0}$: if the swimmer initially 
moves along ${\bf e}$, i.e., ${\dot {\bf R}} \parallel {\bf e}$, we have ${\bf F}_B \parallel {\bf e}$; since we also have $({\bf R}-{\bf R}_0) \parallel {\bf e}$, the torque acting on the body sphere
reads ${\bf T}_B=({\bf R}-{\bf R}_0) \times {\bf F}_B=\mathbf{0}$. Analogously, we find ${\bf T}_P=\mathbf{0}$.
This argument holds true in the presence of hydrodynamic interactions at the normal Oseen level since ${\bf u}({\bf R})=G_0({\bf R},{\bf R}+L(t){\bf e})\cdot (-{\bf F}_P)
\propto {\bf e}$.
However, this finding might (and probably is) 
violated by 
correction terms to the Oseen tensor due to viscosity gradients. Following the previous section, these `violation terms' do however create only 
corrections to the velocities which are on the order of $\dot {\bf R} a \lambda/\eta({\bf R})$ and are therefore negligible for small viscosity gradients.

Now using force balance on the swimmer, ${\bf F}_B+{\bf F}_P={\bf 0}$, and the fact that ${\bf e}$ is time independent, we  
obtain (at normal Oseen level)
\begin{widetext}
\1
\dot {\bf R}=\4{-a_P(t) \dot L(t) \left(\eta ({\bf R}) - \4{3 a_B}{2L(t)} \eta({\bf R}+L(t){\bf e}) \right)}
{a_B \eta({\bf R}) + a_P \eta({\bf R}+L(t){\bf e}) - \4{3 a_B a_P(t)}{2L(t)}\left[\eta({\bf r})+\eta({\bf R}+L(t){\bf e}) \right] } {\bf e}. \label{speedode}
\2
\end{widetext}

This is a closed differential equation for the speed of the swimmer. Expanding this equation in $\epsilon=L(t)\lambda/\eta({\bf R})$ 
yields in lowest order of $\epsilon$
\1
{\bf v}_0(t):=\dot {\bf R}=\4{-a_P(t) \dot L(t)\left[1-\4{3 a_B}{2 L(t)}\right]}{a_B +a_P(t) -\4{3 a_B a_P(t)}{L(t)} } {\bf e}. \label{viscov}
\2
Here, the terms proportional to $a_B/L$ represent the (leading order) impact of hydrodynamic interactions. 
Without these terms we find the simple expression
\1
{\bf v}_0(t)=- \4{a_P(t) \dot L(t)}{a_B + a_P(t)}{\bf e}. \label{viscovnohi}
\2
The average speed $v_0=\langle |{\bf v}_0(t)| \rangle_t$ is nonzero if the body shape deformations are nonreciprocal, in agreement with the Scallop theorem.
For the example of harmonic driving laws with $a_P(t)=a_0 + a_1\cos(\omega t)$ and $L(t)=L_0 + L_1 \cos(\omega t + \phi_0)$, the resulting speed is proportional to $L_1 \omega \sin(\phi_0)$, 
i.e., the swimmer requires a phase difference breaking reciprocity (in the sense that the components of ${\bf v}(t)$ do not feature a mirror symmetry point $t=t_0$) to effectively move forward. 
Here, the swimmer moves `forward' (in the direction of its average motion)
during a part of the body oscillation (stroke) and backwards during the remaining part of the oscillation (recovery stroke), in a similar way as \textit{Chlamydomonas} \cite{Guasto2010}. 
Here, for simplicity, we have considered a viscosity-independent oscillation law, leading to a viscosity-independent swimming speed.
However, 
for a swimmer affording a certain power to deform its body, the frequency of the body oscillations would (at some point)
decrease with increasing viscosity. Following (\ref{viscov}) or (\ref{viscovnohi}), such a frequency-decrease would cause a slow-down of the swimmer with increasing viscosity -- in a similar way 
as our simpler model 
in the main text predicts. 

Correction terms due to the viscosity gradient are of order $\lambda L/\eta_0$:
\1
{\bf v}_1(t)= -\4{\lambda L(t)({\bf e})_y \dot L(t)}{\eta_0} \4{a_B a_P(t) \left(4-9 \4{a_B a_P(t)}{L^2(t)}\right)}{4\left(a_B + a_P(t) - \4{3 a_B  a_P(t)}{L(t)}\right)^2} {\bf e}
\2
Here, $({\bf e})_y$ is the $y$-component of ${\bf e}$, which is time-independent as the orientation of the 
swimmer does not change here.
Depending on the driving law, ${\bf v}_1$ may either weakly slow down or speed up the swimmer when moving up the gradient.
For a swimmer which would be subject to noise or other processes changing it's orientation, ${\bf v}_1$ may cause a (very small) effective drift down or up the gradient at long timescales.
Higher order correction terms can be calculated in principle either by numerically solving (\ref{speedode}) or by expanding up to higher orders in $\epsilon$.
\\For nonuniaxial swimmers
moving by body shape deformations the same mechanism which leads to viscoattraction in swimmers moving due to effective forces (main text) should apply of course. 
One simple example closely mimicking the physical picture illustrated in Fig.~2 of the main text, 
could consist of the uniaxial pushmepullyou-like model which we have just discussed, but with additional spheres rigidly connected to the body sphere 
at both sides of the axis 
connecting $(B)$ and $(P)$.

\begin{acknowledgments}
H.L. acknowledges funding from the SPP 1726 of the Deutsche Forschungsgemeinschaft (DFG, German Research Foundation). B.t.H.\ gratefully acknowledges financial support through a Postdoctoral Research Fellowship from the Deutsche Forschungsgemeinschaft -- HA 8020/1-1.
We thank A.\ M.\ Menzel and A.\ N.\ Morozov for useful discussions. 
\end{acknowledgments}

\bibliographystyle{apsrev4-1}
\end{document}